\documentclass[aps,pre,twocolumn,groupedaddress,showpacs,preprintnumbers,amsmath,amssymb,floatfix]{revtex4}
\usepackage{graphicx}
\usepackage{bm}

\begin{document}

\title{Chiral molecule adsorption on helical polymers}

\author{Maria R. D'Orsogna$^{1}$ and Tom Chou$^{2}$}
\affiliation{$^{1}$Department of Physics \& Astronomy, 
University of California, Los Angeles, CA
90095-154}
\affiliation{$^2$Department of Biomathematics, 
University of California, Los Angeles, CA
90095-1766} 

\date{\today}

\begin{abstract}
We present a lattice model for 
helicity induction on an optically inactive
polymer due to the adsorption of exogenous chiral
amine molecules.
The system is mapped onto a one-dimensional
Ising model characterized by an on-site 
polymer helicity variable and an amine occupancy one.
The equilibrium properties are 
analyzed at the limit of strong coupling between
helicity induction and amine adsorption
and that of non-interacting adsorbant molecules.
We discuss our results in view of recent experimental 
results.
\end{abstract}


\maketitle

\makeatletter
\global\@specialpagefalse 
\let\@evenhead\@oddhead
\makeatother      

\par

\section{Introduction}
\label{sec:introduction}
The chiral properties of 
macromolecules are of particular
importance in determining the nature of their 
interactions with living material \cite{testa}. 
Most cellular receptor enzymes show preferential
binding towards one of the two enantiomers of a helical molecule,
the other mirror-image form 
being either irrelevant or noxious to cellular functioning.
Helicity induction and control
have also been extensively studied for the purposes of 
asymmetric synthesis \cite{seyden}, enantiomer separation
\cite{separate} and other
chemosensing, pharmaceutical \cite{drugs}
and material science applications \cite{diodes}.

The helicity of biological macromolecules,
such as DNA and proteins, is due to the inherent
chirality of the basic components.
Besides  polymerizing optically active monomers, 
synthetic chiral polymers may be constructed
by asymmetric polymerization of non-chiral molecules
using a chiral catalyst,
or, as recently reported \cite{yashima, yashima1},
by complexation of an optically inactive polymer
with a chiral amine. 
The achiral polymer used in Ref.\,\cite{yashima1, yashima2} is a polyacetylene
(poly-1), a \textit{non-helical} structure which coils 
into a helix upon acid-base
interaction with right- or left-handed amines
and amino-alcohols both in solution and in film.
It is believed that the induced helicity
is due to the twist irregularity of adjacent
double bonds around the single bond of the
$-$ CO$_2$H 
functional group
of poly-1 which coils upon interaction with the
amine bases. Other achiral 
polyacetylenes show the same response
\cite{yashima2, yashima3} to chiral amines which 
may be used 
to induce a preferential screw sense on achiral polymers.
In the experiments of Ref.\,\cite{yashima1, yashima2}
certain synthetic helical polymers (such as poly-6, poly-7)
are also used to separate racemic mixtures, due to their
high chiral recognition abilities.
These polymers can be used as chiral stationary phases
in high performance liquid chromatography (HPLC).

\begin{figure}
\includegraphics[height = 1.8 in]{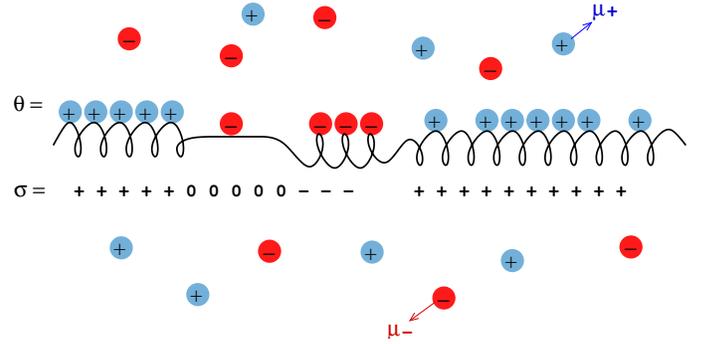}
\caption{
A schematic of a homo-polymer that can acquire three distinct
local
conformations.  For example, the polymer may acquire right or left-handed
helicity, or may remain non-helical. The left region of this figure
represents a
very inducible polymer with its helicity determined by the binding of the
chiral
(labeled by $+$ or $-$) molecules. 
The right branch of the polymer
represents a
locally chiral polymer (which favors say, the $+$ helicity, due to its
intrinsic
structure) that preferentially adsorbs $+$ molecules. These fixed-helicity
polymers have been used to separate $+$ and $-$ chiral molecules in
chromatography.}
\label{draw}
\end{figure}

In this paper, we use a lattice model 
to study helicity induction on a polymer interacting with
adsorbed exogenous chiral molecules
and for chiral discrimination of racemic compounds.
Figure \ref{draw} shows the features of 
the 
one-dimensional lattice model. 
Each monomer $i$ of the polymer is associated to
a pseudo-spin $\sigma_i$ where ${\sigma_i= \left\{-1, \, 0, \, 1\right\}}$
represents the induced left ($\sigma_i = -1$)
or right ($\sigma_i = +1$) monomer helicity.
We also include the possibility for the polymer to 
be uncoiled, in which case $\sigma_i=0$.
Site $i$
is also characterized by the chiral molecule occupancy 
$\theta_i$. If the site is unoccupied 
$\theta_i=0$, if it is occupied by a right (left)-handed molecule
$\theta_i =+ 1 \,   (-1)$. 
The free energy of the polymer $\Omega$
will be modeled using:

\begin{eqnarray}
\Omega \, / k_B T 
& =& - J \sum_i \sigma_i \sigma_{i+1} - \sum_i (h_0 + h_1 \theta_i)
\, \sigma_i  \\
\nonumber
&& - J_2 \sum_i \theta_i \theta_{i+1} 
- \mu_+
\sum_i \delta_{\theta_i,1}
-\mu_- \sum_i \delta_{\theta_i,-1},
\label{ising}
\end{eqnarray}

\noindent
where all parameters appearing on the right-hand side are unitless.
In the Ising model defined by Eq.\,(\ref{ising}),
the ``magnetic field'' $h_0$ 
describes the proclivity of the bare polymer to 
coil into a left or right helix, and
can be interpreted as a net local
torsion of the polymer.
The $h_1$ parameter is
the coupling between the chiral amine
and the polymer.
The $J$ and $J_2$ parameters are the 
nearest neighbor energetic
coupling for the helicity of the
polymer and of the adsorbate respectively.
Field force calculations have been used to estimate
$J$ and $h_0$ for co-polymers of enantiomers \cite{lipson}.

The chemical potentials 
$\mu_\pm$ are 
related to the energy cost
for the adsorption of right or left handed
amines on the polymer, with the delta functions
tracing the number of adsorbed amines.
The last two terms of Eq.\,(\ref{ising})
represent the free energy
contribution arising from polymer-adsorbate
interactions
prior to the coiling of the polymer. 
The chemical potential terms 
$\mu_{\pm}$
are defined so that $h_i \theta_i \sigma_i$
is the energetic split after adsorption of
a molecule with chirality $\theta_i$ on site $i$.

For simplicity we assume that the binding sites and the
basic polymer helical unit are size-commensurate.
Although the microscopic details are more complicated
than our model, they should not affect overall 
equilibrium long wavelength results.
Other than chiral interactions,
we neglect all other effects that the binding 
amines might have on the polymer, such as 
its stiffness although it is well known that binding 
of small molecules affects the persistence length
of polymer chains \cite{diamant}.

Our model is a generalization of a simpler one
in which a random chiral
copolymer with enantiomeric pendant groups 
is described in terms 
of a quenched random-field Ising model \cite{selinger, selinger1, selinger2}.
In our model  $\theta_i$
is an unquenched thermodynamic variable 
related to its chemical potentials $\mu_{\pm}$.
Our model approaches those in Ref.
\cite{selinger, selinger1, selinger2} in the limit
$\{ \sigma_i = +1, -1 \}$, $\{ \theta_i=+1,-1 \}$
and by setting $h_0, J_2, \mu_\pm =0$.

The equilibrium properties of the model 
of  Eq.\,(\ref {ising}) can be evaluated through
the partition function associated with the nine-dimensional
product space $| \left. \sigma_i, \theta_i \right> =
| \left. -,- \right>, \,  | \left. -,0 \right>, \, | \left. -,+ \right>, 
| \left. 0,- \right>, \, 
| \left. 0,0 \right>,  \, | \left. 0,+ \right>, 
\, | \left. +,- \right>, \\
| \left. +,0 \right>,  \, | \left. +,+ \right>$.
In the case of a polymer of length $N$ with 
periodic boundary conditions $\sigma_1 = \sigma_{N+1}$
and $\theta_1 = \theta_{N+1}$, the partition function reads:

\begin{eqnarray}
\Xi  &=& \sum_{\{\sigma_i\}}  \sum_{\{\theta_i\}} e^{-\Omega / k_B T} = 
\mbox{Tr} \, {\bf L}^N , 
\end{eqnarray}

\noindent
where ${\bf L}$ is the nine-dimensional transfer matrix derived from 
$\Omega$.
Thermodynamic quantities such as the free energy $F$,
the average spin-helicity $\left< \sigma_i \right>$, 
the correlation function 
$\Gamma(\ell) = \left< \sigma_i \sigma_{i+\ell} \right> - \left< \sigma_i \right>^2$, 
and the fraction of adsorbed left and right
amines $\left< n_{\theta = \pm 1} \right>$ are obtained from:

\begin{equation}
F = -\ln \Xi, 
\end{equation}

\begin{equation}
N \left < \sigma_i \right>  =  \left(
\frac{\partial \ln \Xi }{\partial 
h_{0}}\right), 
\label{mean}
\end{equation}

\begin{equation}
\left < \sigma_i \sigma_{i+\ell} \right>  = 
\frac{\mbox{Tr} \, \left< \sigma \, {\bf L}^l \, \sigma \, {\bf L}^{N-l} 
\right>}
{\Xi}, 
\label{corr}
\end{equation}

\begin{equation}
\label{nteta}
\text{and} \, \,  \left< n_{\theta = \pm 1} \right> = 
\left< \delta_{\theta_{i},\pm 1}\right >  =  
\left(\frac {\partial \ln \Xi}{\partial \mu_{\pm}}\right).
\end{equation}

\noindent
The $\sigma$ operator is defined as
$\left. \sigma \, | \pm , \theta_i \right> = 
\pm \, \left. | \pm , \theta_i  \right>$, and 
$\left. \sigma | 0, \, \theta_i \right> = 0$. 
Similarly, the $\theta$ operator
is $\left. \theta \, | \sigma_i, \pm \right> = \pm \, \left. 
| \sigma_i, \pm \right>$, and 
$\left. \theta \, | \sigma_i, 0 \right> = 0$.

In the $N \rightarrow \infty$ limit 
Eqs.\,(\ref{mean}-\ref{nteta}) 
can be expressed in terms of the eigenvalues
of ${\bf L}$ ($\lambda_i = \lambda_0 \lambda_1, \dots \lambda_8$) and 
their respective ortho-normalized eigenvectors 
$\left. | \psi_i \right>$. If the largest eigenvalue is
$\lambda_0$, with associated eigenvector $\psi_0$, then \cite{parisi}:

\begin{equation}
\left < \sigma_i \right>  
= \frac {1} {\lambda_0} 
\frac {\partial \lambda_0}{ \partial h_{0}}, 
\label{mean2}
\end{equation}

\begin{equation}
\left < \sigma_i \sigma_{i+\ell} \right>  = \frac{
\left< \psi_0 \, | \,  \sigma \, {\bf L}^{\ell} \, \sigma \, | 
 \, \psi_0 \right>}
{\lambda_0^{\ell}}
\label{corr2}
\end{equation}

\begin{equation}
\label{nteta2}
\left< n_{\theta = \pm 1} \right> = 
\left< \delta_{\theta_{i},\pm 1}\right >  = 
\frac {1} {\lambda_0}
\frac {\partial \lambda_0}{\partial \mu_{\pm}}.
\end{equation}   

\noindent
The amine helicity correlation function
$\left< \theta_i \theta_{i+\ell}\right>$ 
and the polymer-amine correlation function
$\left< \sigma_i \theta_{i+\ell}\right>$ 
can be obtained by substituting the 
appropriate $\theta$ operator
in Eqs.\,(\ref{corr}) and \,(\ref{corr2})
respectively.
The correlation lengths are the same in all cases
and we treat only the polymer-polymer correlation function.
Numerical estimates of the relevant quantities
can be readily computed by diagonalizing the nine-dimensional
transfer matrix.
However it is revealing to explore certain 
physically relevant 
limits analytically. 
In the following sections, we will consider
the experimentally relevant cases of helical polymers
used as chiral discriminants, where $\sigma_i = +1$
for all $i$ sites, and the transfer matrix is three dimensional.
The other physically relevant case is that of
chiral amines used to induce helicity on an optically
inactive polymer. Complexation with left handed
amines yields right handed polymer helicity
so we will consider the four-dimensional 
space given by $\{ \sigma_i = 0, \, +1 \}$ and $\{
\theta_i= -1, \, 0 \}$.
Both these cases shall be analyzed in the limit of
high amine-polymer interaction $(|h_1| \gg 1),$
and of non-interacting adsorbants $J_2 =0$.
We shall also consider the case of a helical polymer
coiling in both screw senses and
interacting with both left and right amine isomers. 
In this case $ \{ \sigma_i = -1, \, +1 \}$.
As above, we shall consider the
limits of strong coupling between
amine occupancy and polymer helicity 
$(h_1 \gg 1),$ 
and the limit
of non-interacting adsorbant amines,
for which $J_2 =0$.

\section{Chiral discrimination}
\label{sec:discrimination}

\noindent
In this section we consider a helical polymer that is chemically fixed
to be right-handed ($\sigma_i = + 1$) and analyze the mean 
fraction of left-handed and right-handed 
chiral amine adsorbates. 
This case corresponds to the parameters $J \rightarrow \infty$
and $h_0 \rightarrow \infty$ and we need only consider the reduced basis
set $\left. |\sigma_i,  \theta_i \right> =
\left. |+,- \right>, \left. |+,0 \right>, 
\left. | +,+ \right>$.
We wish to calculate
the ratio of the two different adsorbate enantiomers:
$r = \left< n_{\theta = -1}\right> /  
\left< n_{\theta = +1}\right>$.
The three-dimensional 
transfer matrix is ${\bf L}$
given by:

\begin{eqnarray}
\nonumber \\
x_+ ^{-1} \, {\bf L} = \left(\begin{array}{ccc}
\frac {y f_- } z & 
1 & \frac {z f_+ }y \\[13pt]
  \frac {f_- }{z} & 
1 & z f_+ \\[13pt]
\frac {f_-}
{y z} & 1  & yz f_{+} 
\end{array}\right).
\nonumber \\
\nonumber \\
\nonumber
\label{L1}
\end{eqnarray}

\noindent
Here, 
$x_{\pm} \equiv e^{(J\pm h_{0}) }, 
y \equiv e^{J_{2}}, 
z \equiv e^{h_{1}}$, and 
$f_{\pm} \equiv e^{\mu_{\pm}}$. 
The largest eigenvalue is readily determined by means of
singular perturbation techniques in the limits of strong coupling
between polymers and amines, and exactly in the case of
non-interacting exogenous molecules.

\subsection{Strong Interaction}
\noindent
In the case of strong interactions between amines and polymers
($h_1 \gg 1$ or $z \gg 1$),
the largest eigenvalue is calculated perturbatively
in $z$ to be:

\begin{eqnarray}
\label{largest}
\lambda_0 \, x_+^{-1} &=& 
 f_+ y z + \frac 1 y + \frac {y-1 + f_+ f_-}{z y^3 f_+}
+ \\ 
\nonumber \\
\nonumber
&& \frac{2 + 2 f_- f_+ (y^2 -1) -3y + y^2}
{f_+^2 y^5 z^2} + {\cal O}[z^{-3}].
\end{eqnarray}

\noindent
Here, $\left< n_{\theta=+1}\right> \gg \left< n_{\theta=-1}\right>
$ because the original choice for the helical polymer $\sigma_i = +1$
and positive values for $h_1$  
favor the adsorption of right-handed amines. 
The fraction of unoccupied sites vanishes as $1/z$ and 
the fraction of left-handed adsorbed amines as $1/z^2$.
The concentration of empty sites, and
the ratio of adsorbed amines is given by:

\begin{eqnarray}
\label{result1}
\left< n_{\theta=0}\right>_{z \rightarrow \infty} &=&
 \frac 1 {f_+ y^2 z } - \frac {3 -2y}{f_+^2 y^4 z^2} \\
\nonumber \\
\nonumber
&\sim& e^{- (h_1 + \mu_+ + 2 J_2)}.
\\
\nonumber \\
\label{result2}
\left[ \frac {\left< n_{\theta = -1}\right> } 
{\left< n_{\theta = +1}\right>} \right]_{z \rightarrow \infty} &=& 
\frac{f_-}{f_+ z^2 y ^4} \\
\nonumber \\
\nonumber
& \sim &  e^{-(\mu_+ - \mu_- + 2 h_1 + 4 J_2)}. \\
\nonumber 
\end{eqnarray}

\noindent
The average adsorbate chirality, 
$\left< \theta_i \right>$
(related to the intensity of the circular dichroism spectra
of Ref.\cite{yashima})
is given by $\left< n_{\theta=+1} \right> - \left< 
n_{\theta= -1} \right>$.
In the opposite limit of negative $h_1$ and $|h_1| \gg 1$,
the adsorption of left handed amines is favored
on a right handed polymer,
and a similar calculation 
yields the same expression
for $\lambda_0$ as Eq.\,(\ref{largest}) with the replacements
$z \rightarrow z^{-1}$, $f_+ \rightarrow f_-$. 
The ratio of right to left-adsorbed amines 
vanishes as $z^2$.

\subsection{Non-interacting Adsorbates}
\noindent
In the case of non-interacting amines,
($J_2 = 0$ or $y=1$) the quantities of interest are 
calculated exactly by means of Eqs.\,(\ref{mean2})
and (\ref{nteta2}) with the largest eigenvalue $\lambda_0$:

\begin{eqnarray}
\lambda_0 \, x_+^{-1}  =
\left< n_{\theta = 0} \right>_{y=1}^{-1}
&=& \frac {f_- + z + f_+ z^2}{z}  \\
\nonumber \\
\nonumber
& \sim & e ^{-(h_1 - \mu_-) } + 1 + e^{(h_1 + \mu_+)}, 
\label{result3} \\
\nonumber \\
\label{result4}
\left[ \frac {\left< n_{\theta = +1}\right>} 
{\left< n_{\theta = -1}\right>}\right]_{y=1} &=& 
\frac{f_+ z^2}{f_-}\\
\nonumber \\
\nonumber
& \sim & e^{+(\mu_+ + 2h_1 -\mu_-)}.
\end{eqnarray}

\section{Helicity Induction}

\noindent
In this section we consider an achiral polymer 
which coils into a preferred helicity, 
right for instance ($\sigma_i = +1$) upon complexation
with external amines of the opposite chirality ($\theta_i = -1$).
This scenario 
is consistent with the experiments of chiral induction
performed with the optically inactive poly-1 chain
in the presence of
chiral molecules of specified right or left helicity \cite{yashima}.
In these experiments, complexation with left handed 
isomers of the chiral molecules 
induce a right-handed helix on the polymer and vice-versa for 
right-handed amines.
In the basis $ | \left. \sigma_i, \theta_i \right> =
| \left. 0,0 \right>,  | \left. 0,- \right>,
|\left. +,0 \right>, | \left. +,- \right>$,
the transfer matrix is four dimensional:

\begin{eqnarray}
{\bf L} = \left(\begin{array}{cccc}
1 & f_- & t & f_- \, t \, z^{-1} \\[13pt]
1 & f_- \,  y & t & f_- \, t \, y  \, z^{-1} \\[13pt]
1 & f_- & x \, t & f_- \, x \, t \, z^{-1} \\[13pt]
1 & f_- \, y & x \, t & f_ - \,  x \, t \, y \, z^{-1}
\end{array}\right)
\nonumber \\
\nonumber
\label{L2}
\end{eqnarray}

\noindent
where $t \equiv e^{h_0}$,
$x \equiv e^{ J }$, so that  $x \, t = x_+$.
For the free energy of the
bare polymer to be minimum at $\sigma_i =0$,
the achiral conformation, we assume that $J, h_0 < 0$.
To model chiral adsorbates which induce polymer helicity
of opposed chirality, we take $h_1 < 0$.

\subsection{Strong Interaction}
\noindent
For highly inducible polymers ($|h_1| \gg 1$
or $z \rightarrow 0$), 
the largest eigenvalue of ${\bf L}$ is:

\begin{eqnarray}
\lambda_0 &=& f_- x \, t \, y \, z^{-1} + 
\frac{1 + t \, x^2 + f_- y^2}{x \, y} + {\cal O}[z].
\end{eqnarray}

\noindent
We can compute  the fraction of adsorbed amines
$\left< n_{\theta=-1}\right>$  and
the mean induced helicity $\left< \sigma_i \right>$.
Note that since $\sigma_i = \left \{ 0, +1 \right \}$ 
the latter is also the fraction 
of $\left< n_{\sigma = +1} \right>$ sites:

\begin{eqnarray}
\label{result5}
\left< n_{\theta=-1}\right>_{z \rightarrow 0} & =& 1 - \frac{1 + t \, x^2}
{f_- x^2 \,y^2 \, t} \, z \\
\nonumber \\
& \sim & 1 - \frac {1 + e^{ -(2J + h_0)}}
{e^{-(h_1 - \mu_- -2J_2)}}.
\nonumber \\ \nonumber \\
\label{result6}
\left< \sigma_i \right> 
= \left< n_{\sigma = +1}\right>_{z \rightarrow 0} &=&  
1 - \frac{1 + f_- \, y^2}
{f_- x^2 \,y^2 \, t} \, z \\
\nonumber \\
& \sim & 1 - \frac {1 + e^{ -(2J_2 + \mu_-)}}
{e^{-(h_1 - h_0 - 2J)}}.
\nonumber
\end{eqnarray}

\subsection{Non-interacting Adsorbates}
\noindent
For non-interacting adsorbates ($J_2 = 0$ or $y=1$) 
the largest eigenvalue for the 
helical induction matrix ${\bf L}$ is:

\begin{eqnarray}
\label{lzero}
\lambda_{0} &=& \frac 1 2 
\left[ 1 + f_- + t \,x + f_- t \, x z^{-1} \frac{}{} + {\cal C}
\right], \\
\nonumber
\\
\mbox{where \, }
{\cal C}&=& z^{-1}
\left[\left(
z + z f_- + z t \, x + f_- t \, x \right)^2 - \right. \\
&& \nonumber
\left.
4 \, t \, z\, (x-1) (1 + f_-) 
\left(z + f_- \right) 
\right]^{1/2}. 
\end{eqnarray}

\noindent
The mean helicity and fraction of occupied amine sites
are again found explicitly through
Eq.\,(\ref{mean2}) and Eq.\,(\ref{nteta2}):

\begin{eqnarray}
\left< n_{\theta=-1}\right>_{y=1}&=& 
\frac {f_-}{2 \, \lambda_0} 
\left[1 + \frac{t  x}z + 
\frac{z + z f_- + (t x)^ 2}
{z \, {\cal C}} -
\right. \\
\nonumber \\
\nonumber
&& \left.
\frac{(x -2)\, [2f_- t +  f_- (t x z)^2 (1 + z)]}
{z\,  {\cal C}}
\right],
\nonumber \\
\left< n_{\sigma= +1}\right>_{y=1} &=& \frac 1 2 + \frac 
{z \,t \, x -z + (t \, x - z) \, f_-}
{2 \,z \, {\cal C}}.
\end{eqnarray}

\section{Random Copolymers}

\noindent
In this section we analyze the case of a polymer
coiling into either screw direction
$\sigma_i = \left\{ -1, +1 \right\}
$ upon interaction with 
exogenous molecules, $\theta_i = \left\{-1, +1 \right\}$, when
no sites are left unoccupied.
We shall also assume $J_2 =0$ throughout the calculation:
this case can be viewed as a dynamic version of a model for
the the assembly of random copolymers \cite{selinger} 
where,
instead of a 
random field $h_i = \left\{+1, -1 \right \}$, assigned with 
probability $p$ and $1-p$ to each site, we introduce the annealed
variable $\theta_i$ whose chemical potentials $\mu_+$ and
$\mu_-$ control the adsorption of each left or right species.
In particular, we wish to compare 
our results with the random field Ising
model of \cite{selinger1}, due the dynamic nature of the
assembly process.
The transfer matrix ${\bf L}$ is defined in the four
dimensional space $\left. | \sigma_i, \theta_i \right>
= \left. | -,- \right>, \left. | -, + 
\right>, | \left.  +, - \right> , 
\left. |+,+ \right> $:

\begin{eqnarray}
{\bf L} = \left(\begin{array}{cccc}
z f_- x_-  & \frac {f_+ x_-} {z} & \frac {f_-} {z x_- } 
& \frac {z f_+} {x_-} \\[13pt]
z f_- x_-  & \frac{f_+ x_-} z \, & 
\frac{f_-} {z x_- }& \frac {z f_+}{x_-}
\\[13pt]
\frac{z f_-}{x_+} & 
\frac {f_+}{z x_+ } & 
\frac {f_- x_+ }{z} & 
z f_+ x_+  \\[13pt]
\frac {z f_-}{x_+}& \frac {f_+}{z x_+} & \frac {f_-  x_+} z & 
z f_+ x_+  \\
\end{array}\right)
\nonumber \\
\nonumber
\label{L3}
\end{eqnarray}

\noindent
and its largest eigenvalues are:

\begin{eqnarray}
\lambda_{0,1} 
&=& \frac {x_- (f_+ + z^2 f_-) + x_+(f_- + z^2 f_+)}{2z} \pm
 \frac 1 {2 x_- x_+ z^2} \\
\nonumber \\
\nonumber
&&
\left \{x_- x_+ [ x_- (f_+ + z^2 f_-) + x_+ (f_- +z^2 f_+)]^2 
\right.+ \\
\nonumber \\
\nonumber
&& \left.
4 x_+ x_- z^2
(1 - x_-^2 x_+^2)(f_+ + z^2 f_-) (f_- + z^2 f_+) \right\}
^{1/2}.
\end{eqnarray}

\begin{figure}
\includegraphics[height = 3.2 in, angle = -90]{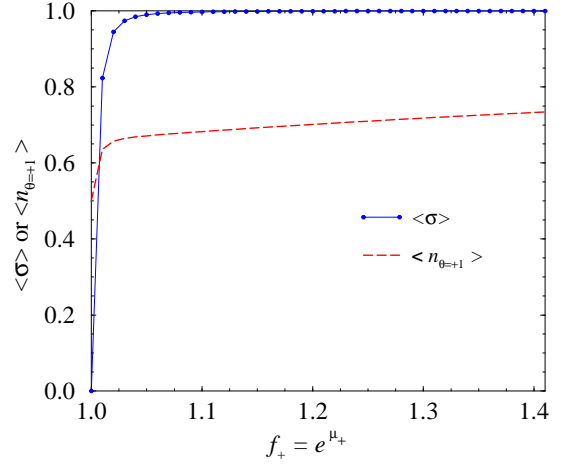}
\caption{Mean helicity and amine occupancy
as a function of $f_+$. The other parameters are fixed at
$x_+ = x_- = 30$ and $h_1 = 1.4$, as in Ref.\cite{selinger} 
and $f_-=1$.
Note the dramatic increase in $\left< \sigma_i \right >$ 
even for values of $f_+$ slightly greater than unity.}
\label{sel}
\end{figure}

\noindent
The mean helicity $\left< \sigma_i \right>$
and fraction of adsorbed amines $\left< n_{\theta = +1} \right>$
can be evaluated by performing the
appropriate derivatives and the correlation length is given 
by Eq.\,(\ref{xi}). Instead of explicitly writing the
expressions for these quantities we plot
$\left< \sigma_i \right>$,  $\left< n_{\theta = +1} \right>$
as a function of $f_+$ for the chosen parameters
$x_+ = x_- = x = 30$, which implies $h_0 =0$,
and $h_1=1.4 $. 
We also set the reference energy at $f_-=1$,
or $\mu_-=0$.
The values for $x$ and $h_1$
correspond to the molecular modeling
values used in \cite{selinger1}. The physical quantities
$\left< \sigma_i \right>$
and $\left< n_{\theta = +1} \right>$
are shown in Figure \ref{sel} as a function
of $f_+$.
As in the case of the Ising random field \cite{selinger1},
a small increase of the fraction of adsorbed amine
$\left< n_{\theta =+1} \right>$ for $f_+ \gtrsim 1$
results in a sharp rise in the optical activity of the sample
as determined by $\left< \sigma_i \right>$,
similarly to the experimental results of \cite{green}.

\section{Chiral Adsorbates and Polymers}
\label{sec:occ-hel}

\noindent
We now analyze the case of a chiral polymer interacting with 
a racemic mixture of chiral adsorbates.
The transfer matrix $\bf L$
determined from Eq.\,(\ref{ising})
in the six-dimensional
space defined by
$ \left. | \sigma_i, \theta_i \right> =
| \left. -,- \right>,   
| \left. -,0 \right>, | \left. -,+ \right>,
| \left. +,- \right>,   
| \left. +,0 \right>, | \left. +,+ \right>$ is:

\begin{eqnarray}
{\bf L} = \left(\begin{array}{cccccc}
yz f_{-}x_{-} & x_{-} & \frac{f_+ x_{-}}{y z} 
& \frac{y f_- }{z x_{-}} & \frac 1 {x_{-}} & \frac{z f_+}
{y x_{-}} \\[13pt]
z f_{-} x_{-} & x_{-} & \frac {f_+ x_{-}} z 
&  \frac{f_-}{z x_{-}} & 
\frac 1 {x_{-}} 
& \frac{z f_{+}}{x_-} \\[13pt]
\frac {z f_{-} x_{-}} y & x_{-} & \frac
{y f_+ x_{-}}{z} & \frac {f_-}{yz x_{-}} 
& \frac 1 {x_{-}} & \frac {yz f_+}{x_-} \\[13pt]
\frac {yz f_- }{x_+} & 
\frac 1 {x_{+}} 
& \frac {f_+} {yz x_{+}} & 
\frac {y f_- x_{+}} z & 
x_{+} & \frac {z f_+ x_+}y \\[13pt]
\frac {z f_{-}}{x_+} & \frac 1 {x_{+}} & 
\frac{f_+}{z x_+} & \frac {f_- x_+}{z} & 
x_{+} & z f_+ x_+ \\[13pt]
\frac{z f_{-}}{y x_+} & \frac 1 {x_{+}} & \frac {y f_+}
{z x_{+}} & 
\frac {f_- x_{+}}
{y z} & x_{+} & yz f_{+}x_{+} 
\end{array}\right)
\nonumber \\
\nonumber
\label{L}
\end{eqnarray}

\noindent 
As above, we shall consider the
limits of strong coupling between
amine occupancy and polymer helicity 
$(|h_1| \gg k_B T),$ or
$z \rightarrow \infty$, or $z \rightarrow 0$ 
and the limit
of non-interacting adsorbant amines,
for which $J_2 =0$, or
$ y=1$.

\subsection{High inducibility}

\noindent
In this section we consider the limit of  $|h_1| \gg 1$,
corresponding to 
$z \rightarrow \infty$ or $z \rightarrow 0$.
In the first case homo-chirality between polymer
and adsorbates are favored, in the latter
opposite screw directions are.
Let us assume $z \rightarrow \infty$.
Eigenvalues and 
and eigenvectors are evaluated in terms of $y$, $p$ and $w$ defined as:

\begin{eqnarray}
p = \frac {f_- x_-}{f_+ x_+} &=& 
e^{(\mu_- -  \mu_+-2h_0)}, \\
\nonumber \\  
w = x_+ x_- &=& e^{2J}.
\end{eqnarray}

\noindent
A direct calculation for the first two eigenvectors $\lambda_{0,1}$  
yields:

\begin{eqnarray}
\lambda_{0,1} &=& \frac {zy f_+ x_+} 2 \left[ (1+p) \pm S(p,w,y) \right], \\
\nonumber \\
S(p,w,y) &=& \left[(1-p)^2 + \frac{4p} {w^2 y^4} \right]^{1/2}.
\end{eqnarray}

\noindent
The other eigenvectors vanish in the limit
$z \rightarrow \infty$ and the mean helicity, from 
Eq.\,($\ref{mean2}$), is:

\begin{equation}
\label{mean3}
\left< \sigma_i \right> = \frac {1-p}{S(p,w,y)}. \\
\end{equation} 

\begin{figure}
\includegraphics[height = 3.2 in, angle = -90]{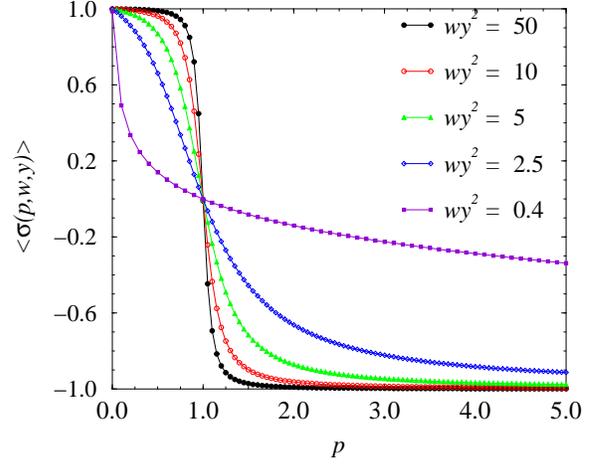}
\caption{Mean helicity 
as a function of $p$ in the case of a polymer with
highly inducible chirality ($h_1 \gg 1$).
In these curves
$ w \, y^2= e^{2(J + J_2)} = 50, 10, 5, 2.5, 0.4$. }
\label{mean1}
\end{figure}

\noindent
Figure \ref{mean1}
shows several $\left< \sigma_i \right>$ curves
for different ferromagnetic and antiferromagnetic values of
$w y^2$ as functions of $p$.
Note that the crossover between positive and negative
$\left< \sigma_i \right>$ values occurs for
$p =1$, which corresponds to
$\mu_+ = \mu_- - 2 h_0$.
In the limit of high helicity-adsorbate coupling,
an overall optically inactive ensemble is reached when 
the balance between the two amine chemical potentials
is offset by the intrinsic torsion $h_0$.
The sharpness of the transition increases with 
the ferromagnetic couplings $J$ and $J_2$
which tend to induce uniform chiralities on adjacent polymer
sites and to adsorb amines with uniform chiralities  
which, in turn, are strongly coupled to the polymer.
The substitution $p \rightarrow p^{-1}$
implies an inversion of the helicity,
and under this transformation 
$\left< \sigma_i \right> \rightarrow 
- \left< \sigma_i \right>$.
The correlation function is defined as
$\Gamma_z(p,w,y,\ell) = \left< \sigma_i \sigma_{i+\ell} \right> -
\left< \sigma_i \right>^2$.
The on-site fluctuation is
$\Gamma_z(p,w,y,\ell=0) = 1 - \left< \sigma_i \right> ^2$ and
in the limit of large distances $\ell$, the correlation function
$\Gamma_z(p,w,x,\ell)$ is determined from Eq.\,(\ref{corr2})
using the fact that $\lambda_n = 0$ for $n > 1$:

\begin{eqnarray}
\label{pro1}
\left< \sigma_i \sigma_{i+\ell}\right> 
&=& \left< \sigma_i \right>^2 + 
\left< \psi_0 | \sigma | \psi_1  \right>
\left< \psi_1 | \sigma | \psi_0  \right>
\left( \frac{\lambda_1}{\lambda_0}\right)^{\ell},
\end{eqnarray}

\noindent
so that:

\begin{eqnarray}
\Gamma_z(p,x,y,\ell \rightarrow \infty) 
&=& D(p,x,y) \, \, e^{-\ell / \xi},
\nonumber \\
\xi^{-1} &=& \ln \left( 
\frac {\lambda_0}{\lambda_1} \right).
\label{xi} 
\end{eqnarray}

\begin{figure}[t]
\includegraphics[height = 3.2 in, angle= -90]{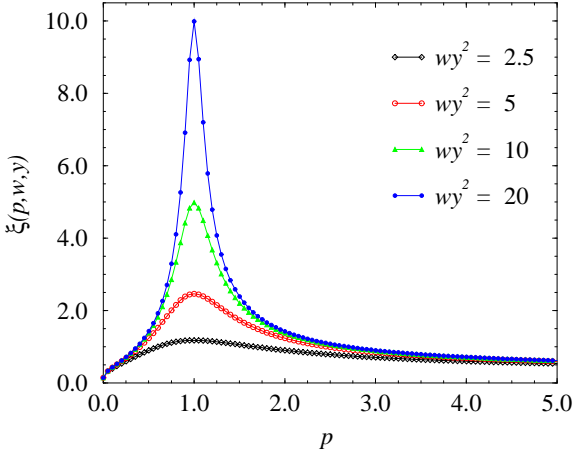}
\caption{
Correlation length $\xi$
as a function of $p$ in the case of a polymer with
highly inducible chirality ($h_1 \gg 1$).
In these curves
$w\, y^2 = e^{2(J + J_2)} = 20, 10, 5, 2.5$.
Note that at the racemic mixture
the correlation length is an increasing function of
$w \,y^2$.}
\label{corro1}
\end{figure}

\noindent
As can be seen from Eq.\,(\ref{mean3}), at 
the antiferromagnetic condition $J + J_2 < 0$
the second non zero eigenvalue $\lambda_1$ is negative,
and in the $z \rightarrow \infty$ limit the
transfer matrix is no longer positive definite.
In calculating the correlation functions we
restrict ourselves to the ferromagnetic case
of $J + J_2 > 0$.
For completeness we estimate the 
prefactor parameter
$D$ using 
the $\left. | \psi_n \right> $ eigenvectors 
found by standard techniques:

\begin{eqnarray}
D(p,x,y) &=& \left [\frac {\lambda_0 \, (1-p+S(p,x,y))}{\lambda_1 \, 
S(p,x,y)}\right]^2 \frac{|\phi_1|^2}{|\phi_0|^2}; \\
\nonumber \\
|\phi_{0,1}|^2 &=& 4p^2 (1+y^2) (w^2 + y^2)
+ \frac {w^4 y^8} 4 \\
\nonumber \\
&&
\nonumber
\left[ 1 + p \pm S(p,x,y)\right]^2 \left[ 1 - p \pm S(p,x,y)\right]^2 + \\
\nonumber \\
&&
\nonumber
16 p w^2 y^4 \left[ 1 - p \pm S(p,x,y)\right]  + w^2 y^4 \\
\nonumber \\
&& 
\nonumber
(1+y^2) (w^2 + y^2)  \left[ 1 - p \pm S(p,x,y)\right]^2 + \\
\nonumber \\
&& 
\nonumber
w^2 y^4  \left[ 1 +p \pm S(p,x,y)\right].
\end{eqnarray}

\noindent
Figure \ref{corro1} 
shows the correlation length
$\xi$ for several values of $w$ and $y$.
The largest $\xi$ occurs at $p=1$ and is :

\begin{equation}
\xi^{-1} (p = 1, w, y) = \ln \, \frac{(w \, y^2 +1)}{ (w \, y^2 -1)}.
\end{equation}

\begin{figure}
\includegraphics[height = 3.2 in, angle = -90]{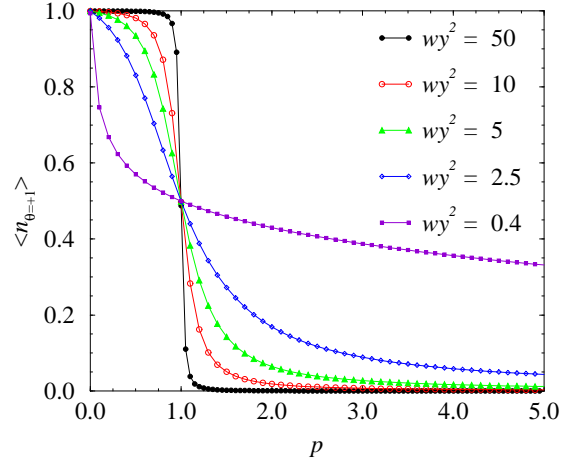}
\caption{Fraction of adsorbed chiral molecules, $n_{\theta_i = 1}$,
for several $w y^2$ values
as a function of $p$ in the strong 
amine occupancy - polymer helicity 
interaction case.
In these plots
,$w y ^2= e ^{2(J + J_2)}= 50, 10 ,5 ,2.5. $}
\label{theta1}
\end{figure}

\noindent
In this limit, $\xi(p=1, w, y)$
is an increasing function of $w \, y^2$:
at higher values of $J+J_2$, large islands of
homo-helicity are formed and the structure is 
increasingly ordered.
The influence of $J_2$ on the mean helicity 
is a consequence of  the  strong interaction between
amines and polymer on-site chirality.
These results are invariant 
under the substitution 
$p \rightarrow p^{-1}$ which signifies an inversion of the
helical screw sense.

We also evaluate the fraction of
adsorbed amine species, $\left< n_{\theta = \pm 1}\right>$.
In the limit of strong coupling
between amine occupancy and polymer helicity,
the two are related by
$\left< n_{\theta = 1}\right> + \left< n_{\theta = -1}\right> = 1$,
as can be seen by direct calculation:

\begin{eqnarray}
\left< n_{\theta = +1} \right> &=& 
\left[ 1-p + \frac{2p}{w^2 y^4} + S(p,w,y) \right] \\
&&\left[ (1-p)^2 + \frac{4p}{w^2 y^4} + S(p,w,y) (1+p) \right]^{-1}.
\nonumber
\end{eqnarray}

\noindent
Figure \ref{theta1} shows the fraction of adsorbed right-handed
chiral molecules  $\left< n_{\theta = +1} \right>$ 
for several values of $w y^2$.
At the optical inactive condition $p=1$, the fraction of adsorbed 
amines is 1/2 for each chirality, 
and all sites are occupied.
In this case, both the average amine and polymer
helicity vanish and the correlation 
function depends on the magnitude of $J$ and $J_2$.
Large values of the latter imply the aggregation of chiral 
amines of the same sign and islands of homo-helicity both for the polymer
and for the adsorbed amines.
These results are again invariant under the substitution
$p \rightarrow p^{-1}$.

\subsection{Non-interacting Adsorbants}
\noindent
Let us now consider the limit $J_2 = 0$ ($y=1$), 
in which the
helicity adsorbant amines do not interact with each other, except for 
exclusion.
A direct calculation of the eigenvalues 
of ${\bf L}$ yields:

\begin{eqnarray}
\lambda_{0,1} &=&\frac {F_+ x_+} {2z} \left[ (1+q) \pm T(q,w)  
\right] \mbox{\, where \, }\\
\nonumber \\
T(q,w) &=& \left[ (1-q)^2 + \frac {4q}{w^2}
\right]^{1/2}, \\
\nonumber \\
F_+ &=& \left[ z^2 f_+ + z + f_- \right] \mbox{\, and} \\
\nonumber \\
q &=& \frac{x_- \left[ z^2 f_- + z + f_+ \right]
}{x_+ \left[ z^2 f_+ + z + f_- \right]} = \frac{x_- F_-}{x_+ F_+}.
\end{eqnarray}

\noindent
The parameters $q$ and $p$ of the previous section
 are related by
$\lim_{ z \rightarrow \infty} q = p$,
and the definition of $w$ is the same as previously defined, 
$w = e^{2J}$.
All other eigenvalues, $\lambda_{n>1}$ are zero.
The mean helicity computed
from Eq.\,(\ref{mean2}) is now:

\begin{eqnarray}
\label{mean4}
\left< \sigma_i \right> &=& \frac {1-q}{T(q,w)}, \\
\nonumber 
\end{eqnarray}

\noindent
which has the same 
form of Eq.\,(\ref{mean3}) upon 
substitution of $p \rightarrow q$ and $y=1$.
In the case of non-interacting amines therefore, 
the non-optically active condition is $q=1$ or,
if we define $2 \mu = \mu_+ + \mu_-$
and $2 \Delta \mu = \mu_+ - \mu_-$:

\begin{equation}
e^{2h_0} = 
\frac {2 + e^ {\mu}
\cosh \left[ \Delta \mu + h_1
\right] }
{2 + e^ {\mu }
\cosh \left[ \Delta \mu - h_1 \right]}.
\end{equation}

\begin{figure}
\includegraphics[height = 3.2 in, angle = -90]{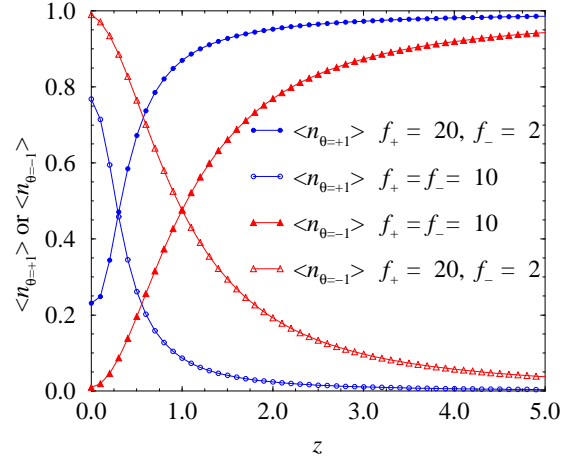}
\caption{Fraction of adsorbed amines for several 
values of $f_{\pm}$
at $h_0 = 3$, $w = 2$ as a function of $z$.
In these plots
the dashed curves correspond to the $\left< n_{\theta = +1}\right> $;
the upper one is at $f_+ = 20, f_- = 2$, and the lower one at 
$f_+ = 10, f_- = 10$.
The solid curves correspond to $\left< n_{\theta = -1}\right> $;
the upper one is at $f_+= 10, f_-=10$ and the lower one at
$f_+ = 20, f_- = 2$.}
\label{ytheta}
\end{figure}

\noindent
In the non-interacting adsorbant limit,
the correlation function $\Gamma_{y=1}(q,w,\ell)$ 
can be evaluated through Eq.\,(\ref{corr2}) 
and it is simply Eq.\,\ref{xi}
with that calculated in 
the strongly interacting limit 
provided $p \rightarrow q$ and $y=1$:

\begin{eqnarray}
\Gamma_{y=1}(q, w, \ell ) = \Gamma_{z} (p, w, y=1, \ell).
\end{eqnarray}

\noindent
The adsorbed amine concentrations $\left<n_{\theta= \pm 1}
\right>$
do not sum to unity and unoccupied sites 
exist. Eq.\,(\ref{nteta2}) yields:

\begin{eqnarray}
\left< n_{\theta=+1}\right> &=& \frac{f_+ z^2}{F_+} +
\displaystyle{
{\cal G} \frac {q f_+}{F_+ F_-}} \, (F_+ -z^2 F_-),  \\
\nonumber \\
\left< n_{\theta=-1}\right> &=& 
\frac{f_- }{F_+} +
\displaystyle{
{\cal G} \frac {q f_-}{F_+ F_-}} (F_+ z^2-  F_-), \\
\nonumber \\
\mbox{where \,} 
{\cal G} &=& \frac{T(q,w) -1 +q + 
\displaystyle{\frac 2 {w^2}}}{T(q,w) \, \left[ \,  T(q,w) + 1 + q \right]}.\\
\nonumber
\end{eqnarray}

\noindent
Several limits are possible for $\left< n_{\theta=0} \right>
=1 - (\left< n_{\theta=+1} \right> + \left< n_{\theta=-1}\right>)$:

\begin{eqnarray}
\left< n_{\theta=0} \right> &=& 1 -\frac z {fz^2 + z + 1}, 
\quad \mbox{when \, } f_+ = f_- \\
\nonumber \\
\left< n_{\theta=0} \right> &=& 1 -\frac z {f_+ + f_- + 1}, 
\quad \mbox{when \, } z=1 \\
\nonumber \\
\left< n_{\theta=0} \right> &=& 1 -\frac {z \cosh ( h_0 )} {e^{h_0} F_+}, 
\quad \mbox{when \, } q=1.
\end{eqnarray}

\section{conclusions}

\noindent
We have constructed
and studied in detail
a generalized lattice theory to
model interactions
between chiral molecules and polymer
chains which may or may not be inherently helical.
One-dimensional Ising models used to describe chiral 
systems have been presented in the literature 
\cite{selinger, selinger1, selinger2} both for 
ordered and quenched helical polymers.
Here, we have introduced the possibility of interactions
with external amines, incorporated as unquenched thermodynamic
variables, as well as the possibility for the
polymer to stay uncoiled, $\sigma_i = 0$.
In particular, for a racemic compound interacting with a
helical polymer, we derived the ratio of the two 
adsorbed enantiomer species:
the results obtained in Eqs.\,(\ref{result1}), 
(\ref{result3}), and (\ref{result4})
can be applied to experimental results to estimate 
microscopic parameters such as the chemical potentials
$\mu_{\pm}$, the amine coupling $J_2$, or the polymer-amine
interaction $h_1$.
Several HPLC experiments estimate the capacity and separation
factors (related to the retention times) of chiral amines interacting
with chiral poly-6 and poly-7 used as a molecular sieve \cite{yashima}. 
However, the retention times arise from the 
adsorption and desorption kinetics of the amines as they 
interact with the stationary phase polymer.
The thermodynamic parameters ($h_1$, $\mu_\pm$ etc.)
are equilibrium properties and by themselves are insufficient
to determine the kinetic capacity factors.
However, if the activation barriers for the amine-polymer
interactions can be independently determined, capacity factors
can then be estimated.

The peak intensity of circular dichroism spectra of 
optically inactive polymers interacting with chiral 
molecules can be directly related to the mean 
induced helicity through Eq.\,(\ref {result5}).
Peak intensity (the mean induced helicity)
increases with larger amine size,  
suggesting
that larger exogenous molecules
result in higher 
amine-amine interaction $J_2$.
This trend is also evident from Eq.\,(\ref{result5}).
Our model does not consider the energetic cost 
to create a left or right handed helicity domain
in an uncoiled region where $\sigma_i=0$.  
An additional free energy term of the form
$ J_1 [\sigma_i^2 + \sigma_i \sigma_{i+1} (\sigma_i + \sigma_{i+1}) +
\sigma_{i+1}^2]$ would be  needed to incorporate such effect.

\vspace{5mm}
\noindent
We thank Prof. J. Rudnick and Z.G. Wang for helpful discussions and we
acknowledge support from the National Science Foundation 
through grant DMS-0206733. TC is also grateful to the Japan 
Society for the Promotion of Science and the National Academy of Sciences
 for the opportunities
provided during the Fourth Annual Symposium on Japanese-American 
Frontiers of Science which motivated this work.




\end{document}